\documentclass[12pt]{article}
\pdfoutput=1
\usepackage{url}
\usepackage{amsmath,amsfonts,mathtools}
\usepackage[final]{graphicx}% Include figure files
\usepackage{color}
\usepackage{lmodern}
\usepackage{verbatim}
\usepackage{bbold}
\usepackage{booktabs}
\usepackage{bm}% bold math
\usepackage{amssymb}
\usepackage{ae}
\usepackage{aecompl}
\usepackage[rftl]{floatflt}
\usepackage[margin=2em, font={small},labelfont={sc}]{caption}
\usepackage{vwcol} % for multiple columnsverbatim
\usepackage{multirow,bigdelim,dcolumn}
\allowdisplaybreaks  % allows pagebreak of amsmath equations
\newcommand{\be}{\begin{equation}}
\newcommand{\ee}{\end{equation}}
\newcommand{\bdm}{\begin{displaymath}}
\newcommand{\edm}{\end{displaymath}}
\newcommand{\bea}{\begin{eqnarray}}
\newcommand{\eea}{\end{eqnarray}}
\newcommand{\bpm}{\begin{pmatrix}}
\newcommand{\epm}{\end{pmatrix}}
\newcommand\lsim{\mathrel{\rlap{\lower4pt\hbox{\hskip1pt$\sim$}}
    \raise1pt\hbox{$<$}}}
\newcommand\gsim{\mathrel{\rlap{\lower4pt\hbox{\hskip1pt$\sim$}}
    \raise1pt\hbox{$>$}}}

\newcommand{\TeV}{\, \mathrm{TeV}}
\newcommand{\GeV}{\, \mathrm{GeV}}

\newcommand{\MGUT}{M_{\rm GUT}}
\newcommand*\xbar[1]{%
   \hbox{%
     \vbox{%
       \hrule height 0.5pt % The actual bar
       \kern0.5ex%         % Distance between bar and symbol
       \hbox{%
         \kern-0.1em%      % Shortening on the left side
         \ensuremath{#1}%
         \kern-0.1em%      % Shortening on the right side
       }%
     }%
   }%
}

\newcommand{\SU}[1]{\ensuremath{{\text{SU}(#1)}}}

\newcommand{\U}[1]{\ensuremath{{\text{U}(#1)}}}

\newcommand{\irrepbase}[2][0]{\ensuremath{\text{\boldirrep{#2}}}}

\newcommand{\boldirrep}{\textbf}
\newlength{\irrepwidth}
\newlength{\irrepbarthickness}
\newlength{\irrepbarheight}
\newcommand{\irrepbarbase}[1]{%
    \settoheight{\irrepbarheight}{\ensuremath{\text{\boldirrep{#1}}}}%
    \settowidth{\irrepwidth}{\ensuremath{\text{\boldirrep{#1}}}}%
    \makebox[0pt][l]{\ensuremath{\text{\boldirrep{#1}}}}%
    \rule[1.2\irrepbarheight]{\irrepwidth}{\irrepbarthickness}%
}

\def\primes#1#2{\count0=#1 \loop \ifnum\count0>0 \advance\count0 by -1 #2\repeat}
\newcommand{\irrep}[2][0]{\ensuremath{\irrepbase{#2}^{\primes{#1}{\prime}}}}
\newcommand{\irrepbar}[2][0]{\ensuremath{\irrepbarbase{#2}^{\primes{#1}{\prime}}}}
\newcommand{\irrepsub}[3][0]{\ensuremath{\irrep[#1]{#2}_{#3}}}
\newcommand{\irrepbarsub}[3][0]{\ensuremath{\irrepbar[#1]{#2}_{#3}}}%
\newcounter{comment}

{\refstepcounter{comment}%
\begin{quote}
\ttfamily\small$\blacksquare$ \textbf{\underline{Comment} $\sharp$\thecomment:}}%
{\end{quote}}

\usepackage{color}
\definecolor{rosso}{cmyk}{0,1,1,0.4}
\definecolor{rossos}{cmyk}{0,1,1,0.55}
\definecolor{rossoc}{cmyk}{0,1,1,0.2}
\definecolor{blu}{cmyk}{1,1,0,0.3}
\definecolor{blus}{cmyk}{1,1,0,0.6}

\usepackage[dvipsnames]{xcolor} 
\usepackage{lscape}
\usepackage{colortbl}
\usepackage{multirow} % merge rows in table
\usepackage{tabularx}
\usepackage{pbox}
\usepackage{hhline}
\usepackage{epstopdf}

\usepackage{url}
\usepackage{amsmath}
\usepackage[final]{graphicx}% Include figure files
\usepackage{caption}
\usepackage{subcaption}%subfigures
\usepackage{bm}% bold math
\usepackage{amssymb}

\usepackage{times}
\usepackage{a4wide}
\usepackage[final]{graphicx}

\usepackage{authblk} % for authors and affiliations
\usepackage[margin=1em,font={small},labelfont={sc}]{caption}

  \renewenvironment{thebibliography}[1]{%
    \begin{oldthebibliography}{#1}%
      \setlength{\parskip}{0ex}%
      \setlength{\itemsep}{0ex}%
      \small
  }%
  {%
    \end{oldthebibliography}%
  }
\begin{document}

\title{\LARGE \bf BSM Matter\\
providing Neutrino Masses and Gauge Unification\footnote{Based on a lecture at the workshop "On a safe road to quantum gravity with matter", Hvar - Croatia, 11-14 September 2018.} }

\author{Ivica Picek}
\affil{Department of Physics, University of Zagreb,
              Bijeni\v{c}ka c. 32, 10002 Zagreb, Croatia}
\date{}
\maketitle

\begin{abstract}
\noindent
I present several scenarios developed in Zagreb, in which TeV-scale particles belonging to  non-trivial weak-isospin multiplets give rise to neutrino-mass mechanisms different from conventional type I, II and III seesaw models.
Two dim 9 tree-level mechanisms, presented first, provide an appealing testability of their exotic TeV-scale particles at the LHC. These models are not genuine, since their particles also provide competing dim 5 loop contributions. The  loop-models presented next are genuine, without competing tree-level contributions. Among them, the three-loop model involves high-order weak multiplets leading to Landau poles. The one-loop model with scalar triplet as the largest multiplet, in addition to good UV properties, provides the particle set promising for gauge coupling unification. Therefore, it served us as a starting point for a study of \SU{5} embedding of additional particles leading to viable unification scenarios. To distinguish among them begs for additional principle which reigns over particle completion and eventual dark matter considerations.
\end{abstract}

\section{Completing the empty places in Periodic Tables}

This year, as International Year of Periodic Table (IYPT 2019), marks the 150th anniversary of the Mendeleev's arranging of the elements in progression of their atomic weights in the columns and the family likeness in the rows of his Periodic Table (left side in Table~\ref{Periodic-Tables}). This arrangement  was followed by two historical subsequent steps.
First, the empty places in table invited a search for the missing elements. The next, revolutionary shift of a focus from atomic weight to atomic number (the number of electrons in its atoms), revealed that the atom is put together by its subatomic pieces. A discovery of the first of them, the electron, by J.J. Thomson in 1897, marks a birth of particle physics as a new science. 

In 2012, the CERN experiments confirmed the spin zero particle corresponding to the long-sought Higgs boson of the Standard Model (SM). In this way a sort of the present day "periodic table" of the subatomic world (right side in Table~\ref{Periodic-Tables}) has been established. At same time, the scalar field column therein is exposed to a program of further Higgs identification, and the neutrino raw seems to offer empty places which invite for an extension of the SM particle content. 
\begin{table}[ht]
\footnotesize
\begin{center}
\begin{tabularx}{\textwidth}{| *{4}{>{\centering\arraybackslash} c|} | *{4}{>{\centering\arraybackslash} c|}}
 \hhline{----||----}
 \multicolumn{4}{|c||}{\raisebox{-0.5ex}{\includegraphics[width=0.45\textwidth]{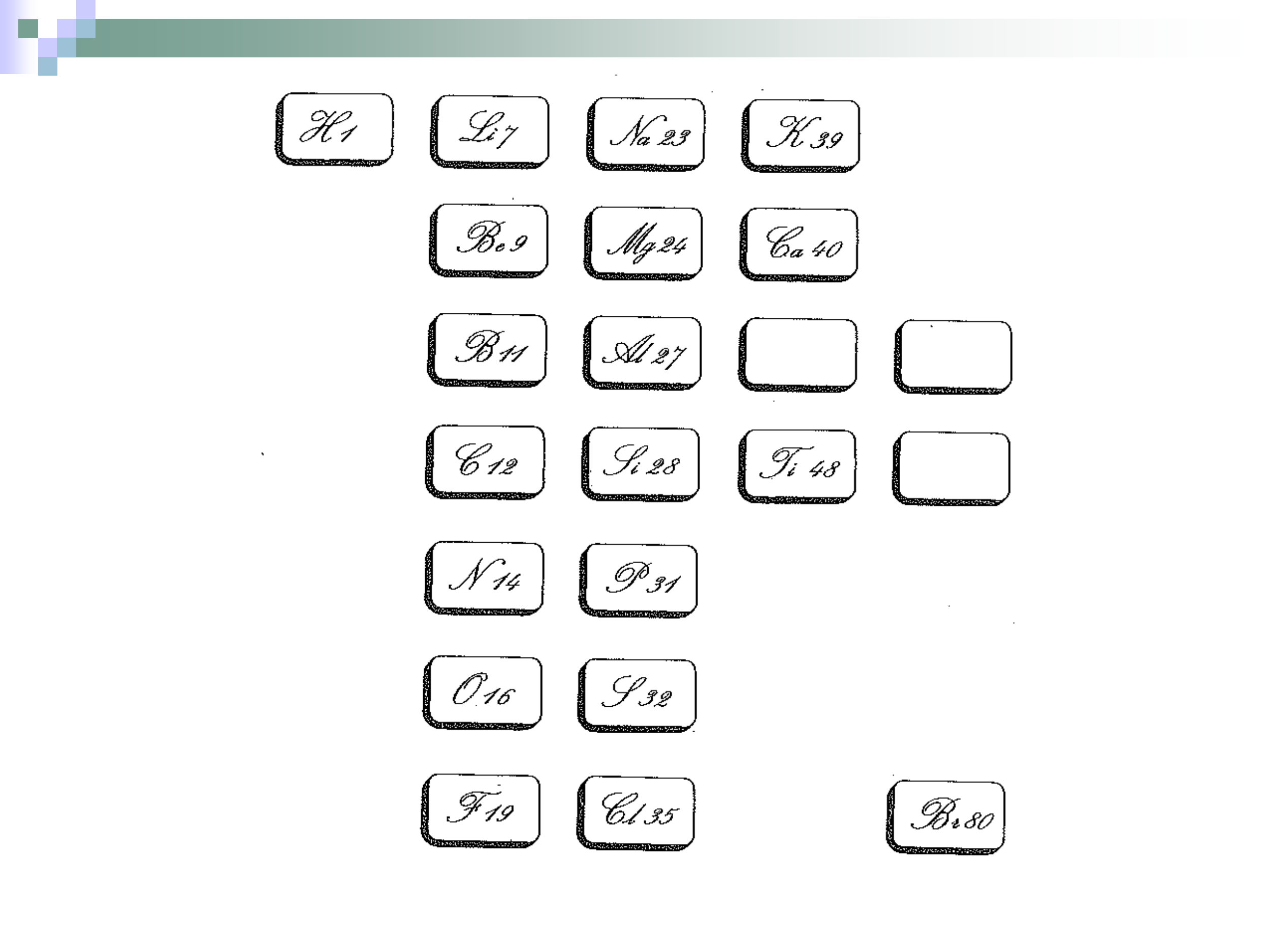}}} & \multicolumn{4}{c|}{\raisebox{-0.5ex}{\includegraphics[width=0.45\textwidth]{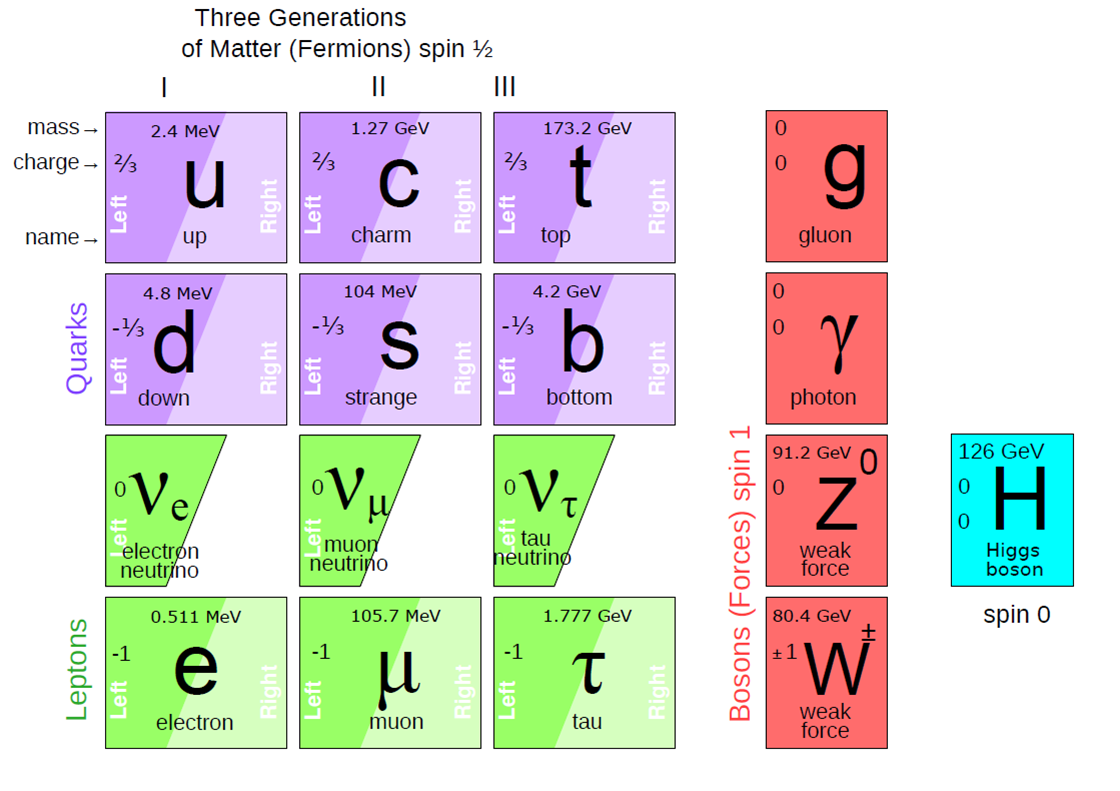}}} \\
 \hhline{----||----}
 \hhline{----||----}
\end{tabularx}
\end{center}
\caption{ \rm  On the left: the Mendeleev's Table of Elements of 1869 (adapted from~\cite{Bronowski:1973} ). On the right: the Standard Model "Periodic Table"(adapted from~\cite{Gninenko:2013tk}). }
\label{Periodic-Tables}
\end{table}
Thereby, the lightness of the neutrinos and the lightness of the Higgs boson stand as prominent landmarks in the landscape of possible SM extensions. In fact, the best motivated new particles appear in attempts to explain small neutrino masses.

In the most popular, seesaw model attempts, the extremely small masses of neutrinos arise on account of their inverse proportionality to large masses of new, yet to be discovered particles. In the simplest, Type I seesaw model~\cite{Minkowski:1977sc-etc}, new heavy particles are singlets under the Standard Model group $\SU{2}_c\times \SU{2}_L\times \U{1}_Y$ which are responsible for active neutrino oscillations~\cite{Gninenko:2013tk}. Still,  devoid of SM charges they are sterile (do not feel SM forces) and resemble the elusive dark matter.
The less elusive seesaw mediators with nontrivial electroweak charges are introduced in the remaining two~\cite{Ma:1998dn} tree-level canonical seesaw mechanisms, dubbed the type II~\cite{KoK-etc} and type III~\cite{Foot:1988aq}.
Their respective scalar triplet and fermion triplet seesaw mediators allow for usual gauge invariant interactions with the SM doublet states. Integrated out, these mediators provide
the Weinberg~\cite{Weinberg:1979sa} dimension five operator  $LLHH$ as the lowest non-renormalizable 
operator that can be constructed with only SM fields. It violates the lepton number by two units, and the observed smallness of the neutrino masses 
can be attributed to to large masses $M \sim 10^{14}$ GeV of the seesaw mediators. The masses of active neutrinos percolate down from the corresponding high scale of lepton number violation (LNV) that may be linked to the point of SM gauge coupling unification hinted first within
\SU{5} Grand Unified Theory (GUT) of Georgi and
Glashow~\cite{Georgi:1974sy}. Notably, the \SU{5} group provides also successful unification of the SM matter, shown in Table~\ref{SMtoSU(5)-Tables}.
\begin{table}[ht]
\footnotesize
\begin{center}
\begin{tabularx}{\textwidth}{| *{4}{>{\centering\arraybackslash} c|} | *{4}{>{\centering\arraybackslash} c|}}
 \hhline{----||----}
 \multicolumn{4}{|c||}{\raisebox{-0.5ex}{\includegraphics[width=0.45\textwidth]{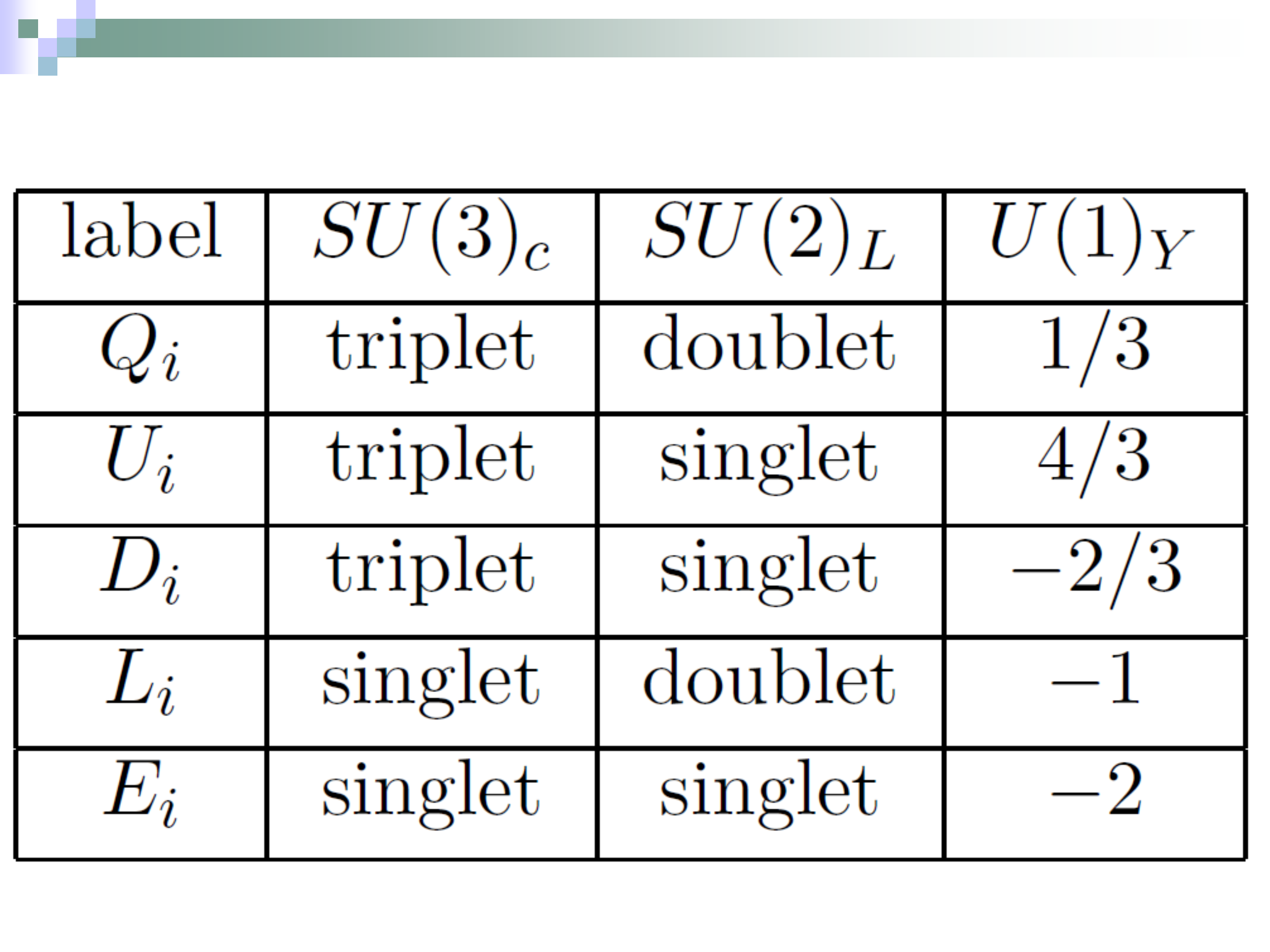}}} & \multicolumn{4}{c|}{\raisebox{-0.5ex}{\includegraphics[width=0.45\textwidth]{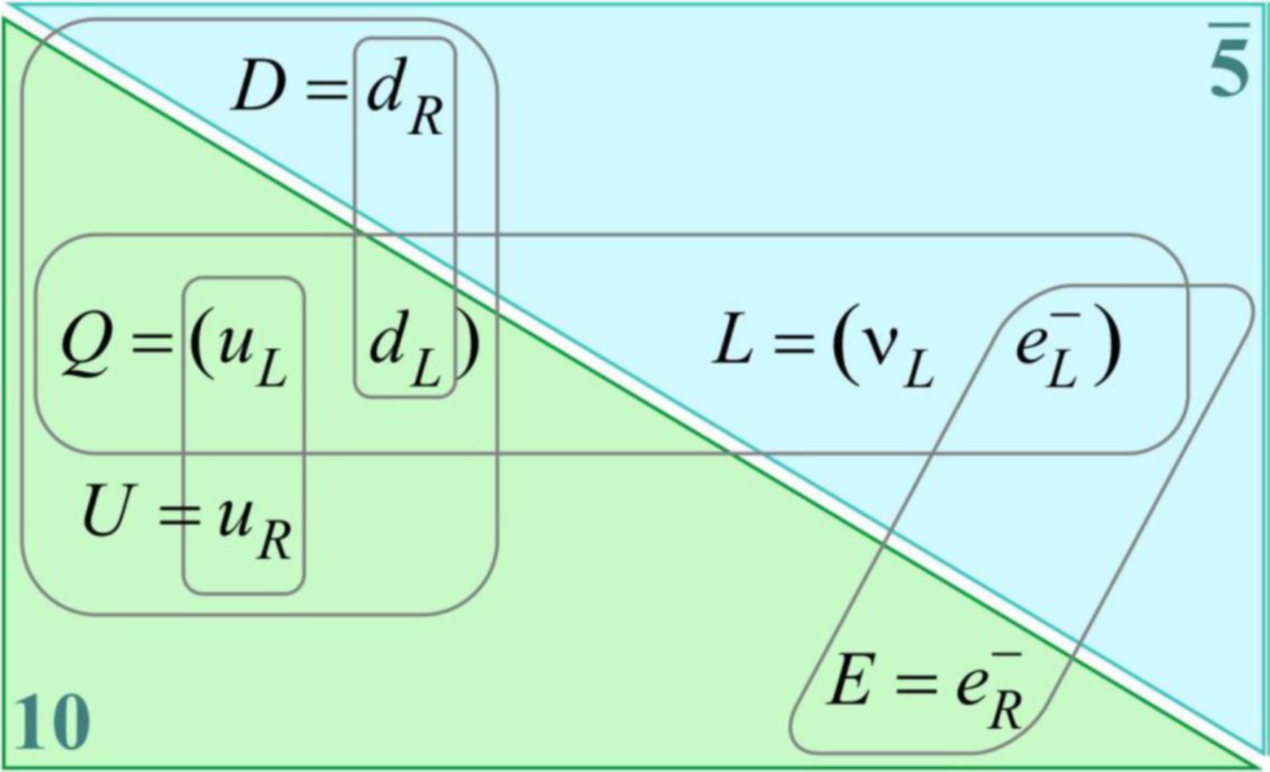}}} \\
 \hhline{----||----}
 \hhline{----||----}
\end{tabularx}
\end{center}
\caption{ \rm  Five species from each family of SM fermions (on the left, adapted from~\cite{Lykken:2006xf}), completing the  sum $\irrepbarsub{5}{F}\oplus \irrepsub{10}{F}$ of two lowest \SU{5} representations (right, adapted from~\cite{C_Smith:2017}). }
\label{SMtoSU(5)-Tables}
\end{table}
 However, in attempts to explain neutrino masses, this SM matter piece would be extended by additional BSM fields, which may open some higher  \SU{5} representations.
 
 The novel tree-level and loop-level neutrino models  proposed in Zagreb rely on TeV-scale particles testable at the LHC. 
Sec. 2 presents two mutually similar dim 9 tree-level neutrino mass models, and Sec. 3 reviews two complementary loop-level models. In Sec. 4 we dwell further on the one-loop model states which satisfy the conditions of perturbativity and vacuum stability, and turn out to be promising for gauge unification.
The viable scenarios of  \SU{5} embedding call for additional colored states, as an additional piece of the UV completion of SM "Periodic Table's" states.
\vspace{.5cm}

%%%%%%%%%%%%%%%%%%%%%%%%%%%%%%%%%%%%%%%%%%%%%%%%%
\section{Two dim 9 Seesaw Models}%%%%%
%%%%%%%%%%%%%%%%%%%%%%%%%%%%%%%%%%%%%%%%%%%%%%%%%

For simplicity, we dub the two models as the "type IV" model employing the custodial scalar quartet fields, and the "type V" model employing the Majorana quintet seesaw mediator. In general, the exotic particles
belonging to higher weak multiplets would be related to dimension $d=5+2n$ operators  $(LLHH)(HH)^n$ studied in \cite{Bonnet:2009ej}. The corresponding
light neutrino mass is given by the  seesaw formula $m_{\nu} \sim v_H (v_H/M)^{d-4}$, where $v_H$= 246 GeV is the vev of the SM higgs.
Accordingly, dimension nine operators have an appeal to involve TeV-scale new particles, which are within the discovery reach of the LHC, and to develop the models at hand.

%%%%%%%%%%%%%%%%%%%%%%%%%%%%%%%%%%%%%%%%%%%%%%%%%
\subsection{Model with Dirac Quintuplet and Custodial Scalar Quartets}
%%%%%%%%%%%%%%%%%%%%%%%%%%%%%%%%%%%%%%%%%%%%%%%%%

The "type IV", dim 9 mechanism, has been introduced in~\cite{Picek:2009is} and further studied in~\cite{Kumericki:2011hf}. We compare its particle content to those of three canonical dim 5 mechanisms in Table~\ref{table_conjunct}. 
\begin{table}[h]
\begin{center}
\begin{tabular}{ccccc}
\hline
Seesaw Type    & Exotic Fermion           & Exotic Scalar                       & Scalar Coupling        & $m_\nu$ at     \\ \hline
Type I         & $N_R\sim(1,0)$           & -                                      & -                     & dim 5          \\
Type II        & -                        & $\Delta\sim(3,2)$                        & $\mu \Delta H H$        & dim 5          \\
Type III       & $N_R\sim(3,0)$           & -                                      & -                     & dim 5          \\ \hline
\textbf{"Type IV"}    & $\Sigma_{L,R}\ (5,2)$  & $(4,-3),\   (4,-1)$   & $\lambda_{1,2}\Phi_{1,2} H H H$   & \textbf{dim 9}          \\\hline
\end{tabular}
\caption{The assignments of electroweak charges for exotic particles introduced by selected neutrino-mass mechanisms.}
\label{table_conjunct}
\end{center}
\end{table}
The three generations of SM leptons $L_L$ and $l_R$ are completed with $n_\Sigma$ vectorlike quintuplets with hypercharge two, $\Sigma_{L,R} \sim (1,5,2)$. Also, the SM Higgs doublet $H$ is completed with two additional scalar quadruplets $\Phi_1$ and $\Phi_2$, transforming as $(1,4,-3)$ and $(1,4,-1)$,
\begin{equation}\label{quintuplet_dir}
\Sigma_{L,R}=\left(\begin{array}{l}
\Sigma^{+++}\\\Sigma^{++}\\\Sigma^{+}\\\Sigma^{0}\\\Sigma^{-}
\end{array}\right)_{L,R}; \ \ \ \
\Phi_1=\left(\begin{array}{l}
\phi_1^0\\\phi_1^-\\\phi_1^{--}\\\phi_1^{---}
\end{array}\right)\ \ , \ \ \ \
\Phi_2=\left(\begin{array}{l}
\phi_2^+\\\phi_2^0\\\phi_2^{-}\\\phi_2 ^{--}
\end{array}\right)\ .
\end{equation}
The scalar potential contains renormalizable terms relevant for our mechanism
\begin{eqnarray}\label{potential_dirac}
V(H, \Phi_1, \Phi_2) &\supset& -\mu_H^2 H^\dagger H + \mu^2_{\Phi_1} \Phi^\dagger_1 \Phi_1+ \mu^2_{\Phi_2} \Phi^\dagger_2 \Phi_2 + \lambda_H (H^\dagger H )^2 \nonumber \\
 &+& \{ \lambda_1 \Phi^*_1 H^* H^* H^* + \mathrm{H.c.} \} + \{ \lambda_2 \Phi^*_2 H H^* H^* + \mathrm{H.c.} \} \nonumber\\
 &+& \{ \lambda_3 \Phi^*_1 \Phi_2 H^* H^* + \mathrm{H.c.} \} \ ,
\end{eqnarray}
where $\lambda_1$ and $\lambda_2$ terms induce vevs for the scalar quadruplets, 
\begin{equation}\label{vev_dir}
    v_{\Phi_1} \simeq -\lambda_1 \frac{v_H^3}{\mu^2_{\Phi_1}} \ \  ,\ \ 
v_{\Phi_2} \simeq -\lambda_2 \frac{v_H^3}{\mu^2_{\Phi_2}} \ \ .
\end{equation}
As pointed out in~\cite{Kumericki:2011hf}, these vevs change the electroweak $\rho$ parameter by  $-6 v_{\Phi_1}^2/v_H^2$ and $-6 v_{\Phi_2}^2/v_H^2$, respectively. Consequently, in case of degenerate vevs $v_{\Phi_1}=v_{\Phi_2}$ these two quartets form a custodial pair, as a generalization of the well studied custodial triplet from the Georgi-Machacek model~\cite{Georgi:1985nv}.\\

%%%%%%%%%%%%%%%%%%%%%%%%%%%%%%%%%%%%%%%%%%%%%%%%%%%%%%%%%%%
{\bf{Neutrino masses}}%%%%%%%%%%%%%%%%%%%%%%%%%%%%%%%
%%%%%%%%%%%%%%%%%%%%%%%%%%%%%%%%%%%%%%%%%%%%%%%%%%%%%%%%%%%

\begin{figure}[h]
\begin{center}
\includegraphics[scale=1.1]{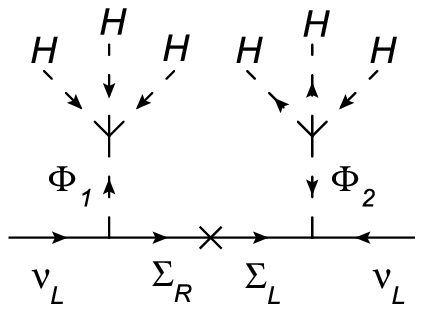} \hspace{0.9cm}
\includegraphics[scale=1.1]{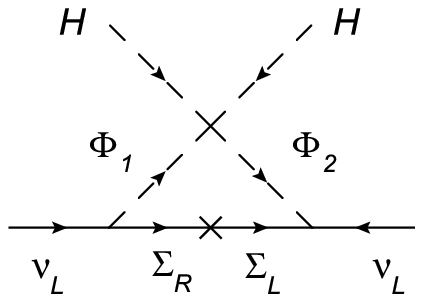}
\caption{Tree-level contribution (LHS) and one-loop contribution (RHS) to neutrino masses.
}
\label{dim9_dir}
\end{center}
\end{figure}

Gauge invariant Lagrangian containing the Yukawa term in
\begin{eqnarray}\label{lagrangian_dirac}
\mathcal{L}&\supset& 
\overline{\Sigma_R}Y_1L_L\Phi_1^* + \overline{(\Sigma_L)^c} Y_2 L_L \Phi_2+ \mathrm{H.c.} 
\ ,
\end{eqnarray}
and the induced vevs $v_{\Phi_1}$ and $v_{\Phi_2}$ lead to the mass terms connecting the SM lepton doublet with new Dirac quintuplet lepton. Three neutral left-handed fields $\nu_L$, $\Sigma_L^{0}$ and $(\Sigma_R^{0})^c$ span the symmetric neutral mass matrix as follows:
\begin{eqnarray}
\mathcal{L}_{\nu \Sigma^0} =  \, -\frac{1}{2}
\left(  \overline{(\nu_L)^c}  \; \overline{(\Sigma_L^0)^c} \; \overline{\Sigma_R^0} \right)
\left( \! \begin{array}{ccc}
0 & m_2^T & m_1^T \\
m_2 & 0 & M_{\Sigma}^T \\
m_1 & M_{\Sigma} & 0
\end{array} \! \right) \,
\left( \!\! \begin{array}{c} \nu_L \\ \Sigma_L^0 \\ (\Sigma_R^0)^c \end{array} \!\! \right)
\; + \mathrm{H.c.}\ .
\label{matrix_neutral_dir}
\end{eqnarray}
The diagonalization of this mass matrix leads to tree-level contribution to the light neutrino mass, shown on 
LHS of Fig.~\ref{dim9_dir}.
In addition, the quartic coupling $\lambda_3$ in Eq.~(\ref{potential_dirac}) gives the loop contribution to the neutrino masses, shown on RHS of Fig.~\ref{dim9_dir}. These two contributions read
\begin{equation}\label{tree_dir}
    m_{\nu}^{tree} \sim \frac {Y_1 Y_2\ \lambda_1\lambda_2\ v_H^6} {M_{\Sigma}\ \mu^2_{\Phi_1}\ \mu^2_{\Phi_2}} \ \ \ \ \ \ , \ \ \ \ \ \ 
    m_{\nu}^{loop} \sim \frac {Y_1 Y_2\ \lambda_3\ v_H^2} {16 \pi^2 \ M_{\Sigma}} \ \ \ ,
\end{equation}
and correspond to dimension nine tree-level seesaw mechanism and to dimension five radiative mechanism, respectively. 
Thereby, for TeV-scale exotic scalars the loop contribution dominates.\\

%%%%%%%%%%%%%%%%%%%%%%%%%%%%%%%%%%%%%%%%%%%%%%%%%%%%%%%%%%%%%%%%%%%%%%%%%%%%%%%%%%%%
{\bf{Production and decays of Dirac quintuplet leptons at the LHC}}%%%%%%%%%%%
%%%%%%%%%%%%%%%%%%%%%%%%%%%%%%%%%%%%%%%%%%%%%%%%%%%%%%%%%%%%%%%%%%%%%%%%%%%%%%%%%%%%

\begin{figure}[h]
\begin{center}
\includegraphics[scale=0.52]{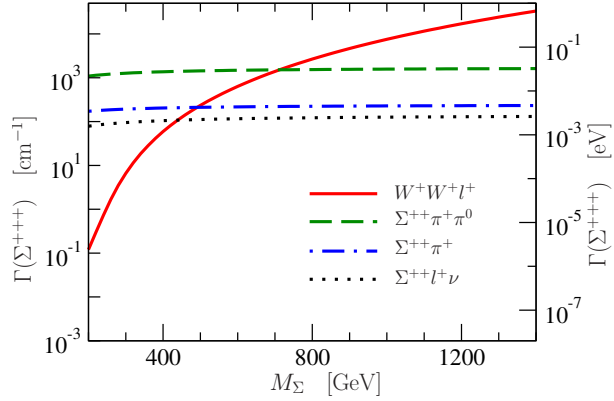}
\caption{Comparison of the partial decay widths of $\Sigma^{+++}$ Dirac quintuplet lepton in dependence of heavy quintuplet mass $M_{\Sigma}$.}
\label{decay+++_dir_800}
\end{center}
\end{figure}

The production channels of the heavy quintuplet leptons in proton-proton collisions are dominated by the quark-antiquark annihilation, $q + \bar{q} \to A \to \Sigma + \bar{\Sigma}\;, \ , (A = \gamma, Z, W^\pm)$, and are determined entirely by gauge couplings of neutral and charged gauge bosons.
  By testing the heavy lepton production 
cross sections one can hope to identify the quantum numbers of Dirac quintuplet particles, but in order to confirm their relation to neutrinos one has to study their decays.
Provided that the exotic scalar states are slightly heavier than the exotic leptons, the exotic scalars will not appear in the final states in  heavy lepton decays. Thereby we distinguish the following classes:\\
%%%%%%%%%%%%%%%%%%%%%%%%%%%%%%%%%%%%%%%%%%%
\textbf{Pointlike decays}%%%%%%%%%%%%%%
%%%%%%%%%%%%%%%%%%%%%%%%%%%%%%%%%%%%%%%%%%%
$\ $to gauge bosons and SM leptons, experienced by four lowest states out of the five $\Sigma$-states properly ordered in Eq.~(\ref{quintuplet_dir}).  Among them, the same-sign dilepton events represent a distinguished signature at the LHC.
On the other hand, the triply-charged $\Sigma^{+++}$ state has other interesting decays: \\
%%%%%%%%%%%%%%%%%%%%%%%%%%%%%%%%%%%%%%%%%%%%%%%%%%
\textbf{Cascade decays}%%%%%%%%%%%%%%%%%%%%%%%
%%%%%%%%%%%%%%%%%%%%%%%%%%%%%%%%%%%%%%%%%%%%%%%%%%
$\ $ $\Sigma^i \to \Sigma^j \pi^+$ and $\Sigma^i \to \Sigma^j l^+\nu$ are suppressed by small mass differences, except for $\Sigma^{+++}$ decays. 
These decays will serve as the referent decays for the  \\
%%%%%%%%%%%%%%%%%%%%%%%%%%%%%%%%%%%%%%%%%%%%%%%%%%%%%%%%%%%%%%%%%%%%%%%%%
\textbf{Golden decay}%%%%%%%%%%%%%%%%%
%%%%%%%%%%%%%%%%%%%%%%%%%%%%%%%%%%%%%%%%%%%%%%%%%%%%%%%%%%%%%%%%%%%%%%%%%
$\ $mode $\Sigma^{+++} \to W^+ W^+ l^+$, having the partial width plotted on Fig.~\ref{decay+++_dir_800}, $\Gamma(\Sigma^{+++} \to W^+ W^+ l^+) \sim M_\Sigma^5 / M_W^4$ in the limit $M_\Sigma \gg M_W$. On the same figure we plot the partial widths for other decays of $\Sigma^{+++}$. A mere nonobservance of triply-charged Dirac fermions may put in a forefront a Majorana quintuplet from our second dim 9 model.

\vspace{.5cm}

%%%%%%%%%%%%%%%%%%%%%%%%%%%%%%%%%%%%%%%%%%%%%%%%%%%%%%%%%%%%%%%
\subsection{Model with Majorana Quintuplets}%%%%%%%%%%%%%%%%%%%%%%
%%%%%%%%%%%%%%%%%%%%%%%%%%%%%%%%%%%%%%%%%%%%%%%%%%%%%%%%%%%%%%

This model~\cite{Kumericki:2012bh} adds to SM fermions three generations of hypercharge zero lepton quintuplets $\Sigma_R=(\Sigma_R^{++},\Sigma_R^{+},\Sigma_R^{0},\Sigma_R^{-},\Sigma_R^{--})$, transforming as $(1,5,0)$ under the SM gauge group. Also, in addition to SM Higgs doublet $H= (H^+, H^0)$ there is a scalar quadruplet $\Phi=(\Phi^{+},\Phi^{0},\Phi^{-},\Phi^{--})$ transforming as $(1,4,-1)$, corresponding to  $\Phi_2$ in Eq.~(\ref{quintuplet_dir})
The gauge invariant and renormalizable Lagrangian
\begin{equation}\label{lagrangian_majorana}
   \mathcal{L} \supset 
   - \big(\overline{L_L} Y \Phi  \Sigma_R + {1 \over 2} \overline{(\Sigma_R)^C} M \Sigma_R + \mathrm{H.c.}\big)
  - V(H,\Phi)  \ .
\end{equation}
contains is the Yukawa-coupling matrix $Y$ and the mass matrix $M$ for two charged Dirac fermions and one neutral Majorana fermion
\begin{equation}\label{fermions_maj}
   \Sigma^{++} = \Sigma^{++}_R + \Sigma^{--C}_R\ ,\ \Sigma^+ = \Sigma^+_R - \Sigma^{-C}_R\ ,\ \Sigma^0
= \Sigma^0_R + \Sigma^{0C}_R\ .
\end{equation}
 Assuming real quartic couplings, the scalar potential has the gauge invariant form
\begin{eqnarray}\label{scalar-pot}
\nonumber  V(H,\Phi) &=& -\mu_H^2 H^\dagger H + \mu_\Phi^2 \Phi^\dagger \Phi + \lambda_1 \big( H^\dagger H \big)^2 + \lambda_2 H^\dagger H \Phi^\dagger \Phi + \lambda_3 H^* H \Phi^* \Phi \\
\nonumber   &+& \big( \lambda_4 H^* H H \Phi+ \mathrm{H.c.} \big) + \big( \lambda_5 H H \Phi \Phi + \mathrm{H.c.} \big) + \big( \lambda_6 H \Phi^* \Phi \Phi+ \mathrm{H.c.} \big)\\
            &+& \lambda_7 \big( \Phi^\dagger \Phi \big)^2 + \lambda_8 \Phi^* \Phi \Phi^* \Phi  \ .
\end{eqnarray}
Here, the presence of the $\lambda_4$ term leads to the induced vev $v_\Phi \sim - \lambda_4 v_H^3 / \mu_\Phi^2$.\\

%%%%%%%%%%%%%%%%%%%%%%%%%%%%%%%%%
{\bf{Neutrino masses}}%%%%%
%%%%%%%%%%%%%%%%%%%%%%%%%%%%%%%%%

\begin{figure}[h]
\begin{center}
\includegraphics[scale=1.1]{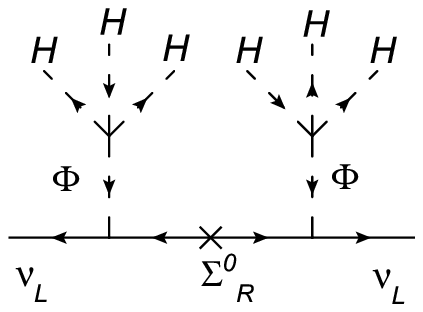} \hspace{0.9cm}
\includegraphics[scale=1.1]{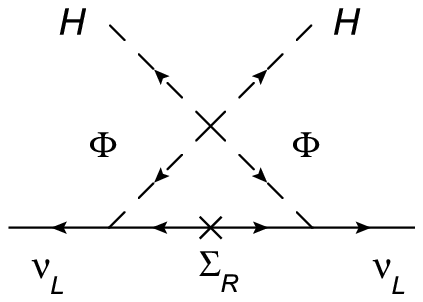}
\caption{\small Tree-level diagram corresponding to dimension-nine operator and one-loop diagrams corresponding to dimension-five operator.}
\label{dim9op}
\end{center}
\end{figure}

\begin{figure}
\centerline{\includegraphics[scale=1.00]{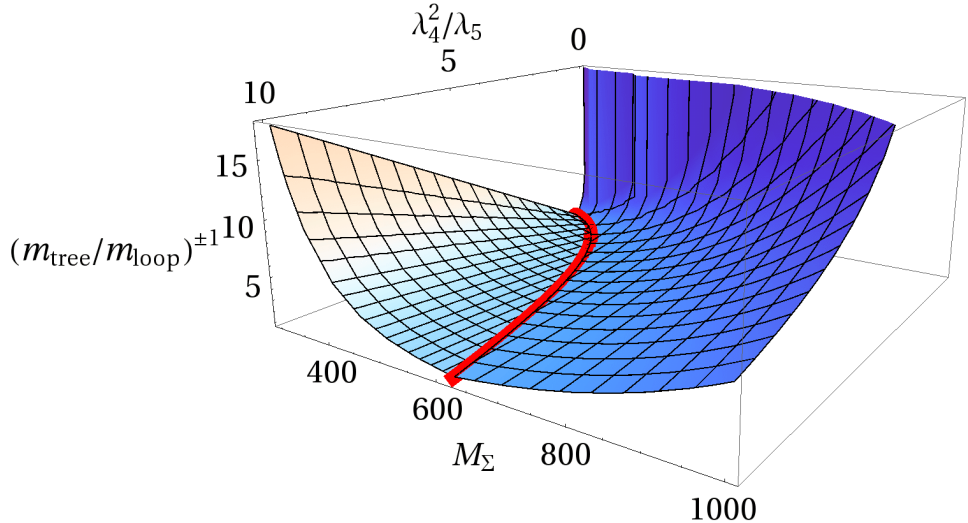}}
\caption{\small For the portion of the parameter space left from the equality-division line, where the tree-level contribution to neutrino masses dominates, we plot $m_{tree}/m_{loop}$. Right from the equality-division line the one-loop contribution to neutrino masses dominates, and we plot $m_{loop}/m_{tree}$.}
\label{tree_dom-ce}
\end{figure}

The Yukawa couplings in Eq.~(\ref{lagrangian_majorana}) and
the vev $v_\Phi$ generate a Dirac mass term connecting $\nu_L$ and $\Sigma_R^0$, a nondiagonal entry in
the mass matrix for neutral leptons given by
\begin{eqnarray}
\mathcal{L}_{\nu \Sigma^0} =  \, -\frac{1}{2}
\left(  \overline{\nu_L}  \; \overline{(\Sigma_R^0)^C} \right)
\left( \! \begin{array}{cc}
0 & \frac{1}{\sqrt{2}} Y v_\Phi \\
\frac{1}{\sqrt{2}} Y^T v_\Phi & M
\end{array} \! \right) \,
\left( \!\! \begin{array}{c} (\nu_L)^c \\ \Sigma_R^0 \end{array} \!\! \right)
\; + \mathrm{H.c.}\ .
\label{neutral_mass_matrix}
\end{eqnarray}
After diagonalizing this mass matrix the light neutrinos acquire the Majorana mass contribution corresponding to tree-level diagram displayed on LHS of Fig.~\ref{dim9op}. Simultaneously, the quartic $\lambda_5$ term in Eq.~(\ref{scalar-pot}) generates the one-loop diagram displayed on RHS of Fig.~\ref{dim9op}.
These two contributions added together give the light neutrino mass matrix
\begin{eqnarray}\label{mnu}
  (m_\nu)_{ij} &=& (m_\nu)_{ij}^{tree}+(m_\nu)_{ij}^{loop} \nonumber\\
               &=& \frac{-1}{6} (\lambda^*_4)^2 \frac{v^6}{\mu_\Phi^4} \sum_k {Y_{ik} Y_{jk} \over M_k}
+ {-5 \lambda_5^* v^2 \over 24 \pi^{2}}
\sum_k {Y_{ik} Y_{jk} M_k \over m_\Phi^{2} - M_k^{2}} \left[
1 - {M_k^{2} \over m_\Phi^{2}-M_k^{2}} \ln {m_\Phi^{2} \over M_k^{2}}  \right] \,. \nonumber\\
\end{eqnarray}
In Fig.~\ref{tree_dom-ce} we distinguish the parts of the parameter space for which the tree-level (loop-level) contribution dominates. Generally, the loop contribution dominates for TeV-scale exotic scalars.

%%%%%%%%%%%%%%%%%%%%%%%%%%%%%%%%%%%%%%%%%%%%%%%%%%%%%%%%%%%%%%%%%%%%%%%%%%%%%%%%%%%%
{\bf{Production and decays of Majorana quintuplet leptons at the LHC}}%%%%%%%%
%%%%%%%%%%%%%%%%%%%%%%%%%%%%%%%%%%%%%%%%%%%%%%%%%%%%%%%%%%%%%%%%%%%%%%%%%%%%%%%%%%%%

The production channels of heavy quintuplet leptons in proton-proton collisions studied in Ref.~\cite{Kumericki:2012bh} are
dominated by the quark-antiquark annihilation via neutral and charged gauge bosons.

In order to confirm the relation of the quintuplet particles to neutrinos, their decays have also been studied in~\cite{Kumericki:2012bh}, including the monitoring Higgs diphoton-decay and Higgs to Z-photon decay, induced by doubly-charged scalars. 

The distinctive decay signatures could come from doubly-charged $\Sigma^{++}$ and $\overline{\Sigma^{++}}$ components of the fermionic quintuplets.
The signals which are good for the discovery correspond to relatively high signal rate and small SM background. Two promising classes of events contain $\Sigma^+$ decaying to $e^+$ or $\mu^+$ lepton and $Z^0 \to (\ell^+ \ell^-, q \bar{q})$ resonance  helping in $\Sigma^+$ identification:
\begin{displaymath}
(i) \qquad p \, p \to \Sigma^+  \, \overline{\Sigma^0} \to (\ell^+Z^0) \, (\ell^+W^-) \;,
\end{displaymath}
the LNV event 
%having 0.7 fb 
with the same-sign dilepton state, which is nonexistent in the SM and thus devoid of the SM background;
\begin{displaymath}
(ii) \qquad p \, p \to \Sigma^{++}  \, \overline{\Sigma^+} \to (\ell^+W^+) \, (\ell^-Z^0) \;,
\end{displaymath}
having relatively high signal rate 
which exceeds the SM background.

Let us note that presented tree-level dim 9, "Type IV" and "Type V" mechanisms are not genuine, since they develop lower dim 5 loop contributions. On the other hand, we will present in next section two {\em genuine} loop mechanisms which are free from competing tree-level contributions, without need to forbid them by some ad hoc symmetry.

\vspace{0.5cm}

%%%%%%%%%%%%%%%%%%%%%%%%%%%%%%%%%%%%%%%%%%%%%%%%%
\section{Two Genuine Loop Models}%%%%%
%%%%%%%%%%%%%%%%%%%%%%%%%%%%%%%%%%%%%%%%%%%%%%%%%

After discovery of a Higgs-like $125$ GeV particle~\cite{Aad:2012tfa,Chatrchyan:2012ufa}, 
the present data allow it to be just a detected part of an extended Standard Model (SM) Higgs sector. Additional Higgs 
candidates have been proposed in extensions of the SM Higgs sector aimed at generating neutrino masses at the loop level.
We study a possible appearance of a Higgs partner in the context of beyond-SM (BSM) fields which appear in two different genuine models of radiative neutrino masses, displayed in Table~\ref{models}:\\ 
{\em The one-loop neutrino mass model}~\cite{Brdar:2013iea} with minimal BSM representations, providing the neutral component of a real scalar field $\Delta$  in the adjoint representation of the $SU(2)_L$ as the Higgs relative;\\ 
{\em The three-loop neutrino mass model}~\cite{Culjak:2015qja} with exotic BSM representations, where the Higgs partner
emerges in the form of the heavy CP-even neutral scalar field in the framework of the two-Higgs-doublet  model (2HDM)~\cite{Branco:2011iw}.\\ 
\begin{table}[ht]
\footnotesize
\begin{center}
\begin{tabularx}{\textwidth}{| *{4}{>{\centering\arraybackslash} c|} | *{4}{>{\centering\arraybackslash} c|}}
 \hhline{----||----}
 \multicolumn{4}{|c||}{\raisebox{-0.5ex}{\includegraphics[width=0.45\textwidth]{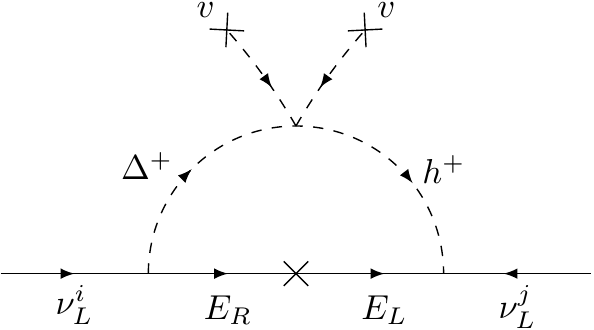}}} & \multicolumn{4}{c|}{\raisebox{-0.5ex}{\includegraphics[width=0.45\textwidth]{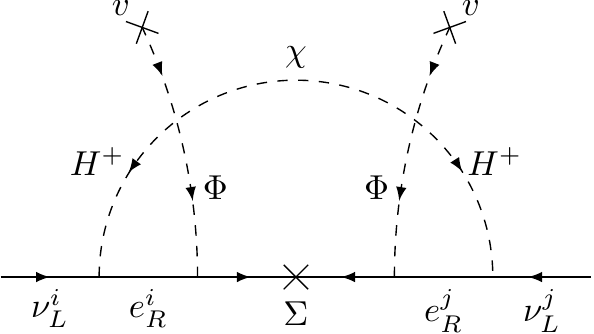}}} \\
 \hhline{----||----}
 \rowcolor[cmyk]{0.2,0.0,0,0.0}  Name &  $SU(2)_L$ &  $U(1)_Y$ & $Q$ &  Name &  $SU(2)_L$ &  $U(1)_Y$ &  $Q$ \\
 \hhline{----||----}
 \rowcolor[cmyk]{0,0,0.2,0} $\textcolor{LimeGreen}{\boldsymbol{\Delta}}$ & $3$ & $0$ & $\pm1,0$ & $\textcolor{LimeGreen}{\boldsymbol{H}}_{\textcolor{LimeGreen}{\boldsymbol{1,2}}}$ & $2$ & $1$  & $0,1$ \\
 \hhline{----||----}
  \rowcolor[cmyk]{0,0,0.2,0} $h^+$ & $1$ & $2$  & $1$ & $ \Phi$ & $5$  & $-2$ &$-3,-2,\pm 1,0$ \\
 \hhline{----||----}
 \rowcolor[cmyk]{0,0.2,0,0.1}$E_{R}$ & $2$ & $-1$ & $0,-1$ & \cellcolor[cmyk]{0,0,0.2,0} $\chi$ & \cellcolor[cmyk]{0,0,0.2,0} $7$ & \cellcolor[cmyk]{0,0,0.2,0} $0$  & \cellcolor[cmyk]{0,0,0.2,0} $\pm3,\pm2,\pm1,0$\\
 \hhline{----||----}
 \rowcolor[cmyk]{0,0.2,0,0.1}$E_{L}$ & $2$ & $-1$  & $0,-1$ & $\Sigma $ & $5$ & $0$ & $\pm2,\pm1,0$\\
 \hhline{----||----}
\end{tabularx}
\end{center}
\caption{ \rm  Neutrino mass models. Scalar fields are in (light) yellow and fermion fields in three generations are in (dark) red. The fields containing an extra Higgs candidate are in (light grey) green. For the one-loop model (left) the SM Higgs doublet  $H_{\rm SM}=(\phi^+,\phi^0)^T$ manifests itself only via its VEV $v$ in the neutrino mass diagram.}
\label{models}
\end{table}
Since the extra scalar states affect the ultra-violet (UV) behavior of a given model, such states cannot come alone or with arbitrary coupling. Previous accounts~\cite{Ferreira:2015rha} and~\cite{Khan:2016sxm} studied a validity up to Planck scale for a sole second Higgs doublet or an additional Higgs triplet, respectively. The present study considers such extra scalars in the setup~\cite{Brdar:2013iea} and~\cite{Culjak:2015qja} decorated with additional fields required to close respective neutrino mass loop diagrams.
The presence of extra vector-like leptons makes both radiative models
to belong to generic setup~\cite{Dermisek:2016via} which enables identification of regions of model parameter space for which the
particular cascade decays may reveal the Higgs partners.

%%%%%%%%%%%%%%%%%%%%%%%%%%%%%%%%%%%%%%%%%%%%%%%%%%%%%%%%%%%%%%
\subsection{The one-loop model} 
%%%%%%%%%%%%%%%%%%%%%%%%%%%%%%%%%%%%%%%%%%%%%%%%%%%%%%%%%%%%%%

The first mass model~\cite{Brdar:2013iea} in our focus is displayed on LHS in Table~\ref{models}. It is a variant of the one-loop model by Zee~\cite{Zee:1980ai}: 
 The charged scalar singlet $h^+ \sim (1,2)$ from Zee's loop-diagram has been kept, while Zee's second Higgs doublet has been replaced by hypercharge zero triplet scalar field $\Delta  \sim (3,0)$
in conjunction with the vector-like lepton $E_{R,L} \sim (2,-1)$ in three generations.
This model has an appeal to invoke low non-singlet weak representations and to be free of imposing additional {\it ad hoc} $Z_2$ symmetry to eliminate the tree-level contribution. This allows a mixing of the triplet scalar field with the SM Higgs field  and its participation in electroweak symmetry breaking (EWSB).
The gauge invariant scalar potential contains new quartic terms
\begin{eqnarray}\label{potential}
    &&V(H_{\rm SM},\Delta,h^+)  \supset
    \lambda_3 (\mathrm{Tr}[\Delta^2])^2 + \lambda_4 H_{\rm SM}^\dag H_{\rm SM} h^-h^+ + \lambda_5 H_{\rm SM}^\dag H_{\rm SM} \mathrm{Tr}[\Delta^2] \nonumber\\
    &&+ \lambda_6 h^-h^+ \mathrm{Tr}[\Delta^2] + (\lambda_7 H_{\rm SM}^\dag\Delta\tilde{H}_{\rm SM}h^+ + \mathrm{H.c.}) + \mu H_{\rm SM}^\dag \Delta H_{\rm SM} \ ,
\end{eqnarray}
and the trilinear $\mu$ term. 
Without imposing $Z_2$ symmetry  the $\mu$ term
leads to an induced VEV  $\langle \Delta^0\rangle$ for the neutral triplet component, which is constrained by electroweak measurements to be smaller than a few GeV.
The neutrino mass matrix obtained from an effective operator displayed in Table~\ref{models} is proportional to $\lambda_7$ coupling in Eq.~(\ref{potential}),
\begin{eqnarray}\nonumber
&&\mathcal{M}_{ij}=\sum_{k=1}^3\frac{[(g_1)_{ik} (g_2)_{jk} + (g_2)_{ik}(g_1)_{jk}]} {8\pi^{2}} \ \lambda_7 \; v_H^2 \; M_{E_k}\\
&&\hspace{1.8cm}
\frac{M_{E_k}^{2}m_{h^+}^{2}\ln{\frac{M_{E_k}^{2}}{m_{h^+}^{2}}}+
M_{E_k}^{2}m_{\Delta^+}^{2}\ln{\frac{m_{\Delta^+}^{2}}{M_{E_k}^{2}}}
+m_{h^+}^{2}m_{\Delta^+}^{2}\ln{\frac{m_{h^+}^{2}}{m_{\Delta^+}^{2}}}}{({m_{h^+}^{2}-m_{\Delta^+}^{2}})
(M_{E_k}^{2}-m_{h^+}^{2})(M_{E_k}^{2}-{m_{\Delta^+}}^{2})} \; .
\label{effective}
\end{eqnarray}
Assuming the mass values  $m_E \sim m_{\Delta^+} \sim m_{h^+} \sim 400$ GeV, the neutrino masses $m_\nu \sim 0.1$ eV can be generated for appropriate Yukawa couplings and for $\lambda_7$ coupling in (\ref{potential}) of the order of $10^{-4}$~\cite{Brdar:2013iea}. 

%%%%%%%%%%%%%%%%%%%%%%%%%%%%%%%%%%%%%%%%%%%%%%%%%%%%%%%%%%%%%%
\subsection{The three-loop model} 
%%%%%%%%%%%%%%%%%%%%%%%%%%%%%%%%%%%%%%%%%%%%%%%%%%%%%%%%%%%%%%
 
The second mass model~\cite{Culjak:2015qja} in our focus is based on the three-loop diagram displayed on the RHS in Table~\ref{models}.
Notably, this model contains 2HDM sector augmented by exotic fields needed to close the three-loop mass diagram: the complex scalar pentuplet $\Phi$ and a real scalar septuplet $\chi$, in conjunction with vector-like lepton $\Sigma_{R,L} \sim (5,0)$. Since $\Phi$ and $\chi$ fields do not form renormalizable gauge invariant couplings with SM fermions, there is no need for an additional symmetry to eliminate the tree-level neutrino mass contributions.
Moreover, the model is fortuitously scotogenic~\cite{Culjak:2015qja}:
a standard discrete $\tilde Z_2$ symmetry imposed to produce a natural flavour conservation in 2HDM results in accidental $Z_2$ odd parity of its BSM sector shown in Table~\ref{Z-charges}.
On account of it, the lightest among the three generations  ($\alpha =1,2,3$) of exotic real fermions $\Sigma_{\alpha} \sim (5,0)$ may be a viable minimal dark matter (MDM)~\cite{Cirelli:2005uq} candidate. 
Out of four different ways the Higgs doublets are 
conventionally assigned charges under a $\tilde Z_2$ symmetry~\cite{Kanemura:2014bqa}, we have adopted 
the  ``lepton-specific" (Type X or Type IV) 2HDM implemented originally in~\cite{Aoki:2008av, Aoki:2009vf}. The ratio of VEVs of 2HDM fields $H_{\bf{1,2}}\sim (2,1)$ is given
by $\mathrm{tan}\beta = v_1/v_2$.
\begin{table}
\begin{center}
  \begin{tabular}{|c|ccccc|cc|ccc|}
   \hline
 \hbox{Symmetry} & $Q_i$ & $u_{i R}$ & $d_{i R}$ & $L_{i L}$ & $e_{i R}$ & $H_{\bf{1}}$ & $H_{\bf{2}}$ & $\Phi$ &
    $\chi$ & $\Sigma_{\alpha}$ \\\hline
\rowcolor[cmyk]{0.1,0,0.1,0}
\cellcolor[cmyk]{0.0,0,0.2,0}$Z_2\frac{}{}$                {\rm accidental} & $+$ & $+$ & $+$ & $+$ & $+$ & $+$ & $+$ & $-$ & $-$ & $-$ \\ \hline  
 \rowcolor[cmyk]{0.2,0.0,0,0.0}
  \cellcolor[cmyk]{0,0.2,0.0,0}
  $\tilde{Z}_2\frac{}{}$ {\rm imposed}& $+$ & $-$ & $-$ & $+$ &
                       $+$ & $+$ & $-$ & $+$ & $-$ & $+$ \\\hline
   \end{tabular}
\end{center}
  \caption{Charge assignment under an automatic $Z_2$ symmetry which is induced by the imposed $\tilde Z_2$ symmetry 
  in the lepton-specific 2HDM.}
  \label{Z-charges}
\end{table}
The fields $H_{\bf{1,2}}$ are expressed in standard way in terms of 
physical charged scalars $H^\pm$ and two CP-even neutral states  $h$ and $H$ which mix with the angle $\alpha$, and are proposed to be the SM-like Higgs $h(125)$ and its heavier relative $H$.

The most general CP-conserving 2HDM potential $V(H_{\bf{1}},H_{\bf{2}})$ presented in~\cite{Culjak:2015qja} is conventionally expressed in terms of five quartic couplings $\lambda_1$ to $\lambda_5$ which can be traded for the four physical Higgs boson masses and the mixing parameter $\sin(\beta-\alpha)$.
The full scalar potential contains additional gauge invariant pieces for additional exotic scalar fields $\Phi \sim (5,-2)$ and $\chi \sim (7,0)$
\begin{eqnarray}\label{scalarpot}
 V(H_{\bf{1}},H_{\bf{2}},\Phi, \chi) &=& V(H_{\bf{1}},H_{\bf{2}}) + V(\Phi)  + V(\chi) + V_m(\Phi,\chi) \\ 
\nonumber   &+&  V_m(H_{\bf{1}},H_{\bf{2}},\Phi) + V_m(H_{\bf{1}},H_{\bf{2}},\chi)
            +  V_m(H_{\bf{1}},H_{\bf{2}},\Phi, \chi)  \ .
\end{eqnarray}
The  EWSB in $\tilde Z_2$-symmetric quartic term 
  $V_m(H_{\bf{1}},H_{\bf{2}},\Phi, \chi) =  \kappa H_{\bf{1}} H_{\bf{2}} \Phi \chi + \mathrm{h.c.}$
leads to the substitution
$\kappa (H^{+}_{\bf{1}} H^{0}_{\bf{2}} +  H^{+}_{\bf{2}} H^{0}_{\bf{1}})  \rightarrow  
v \kappa \cos(2\beta) H^{+}$.
The resulting three-loop diagram, neglecting the mass differences within $\Phi, \chi$ and $\Sigma_{\alpha}$  multiplets, results in the mass matrix $M^\nu_{ij}$ for active neutrinos of the form~\cite{Culjak:2015qja} 
\begin{eqnarray}
M_{ij} &=& \sum_{\alpha=1}^3 
   C_{ij}^\alpha \, F(m_{H^\pm}^{},m_\Phi^{}, m_\chi, m_{\Sigma_{\alpha}}), \label{eq:mij}
\end{eqnarray}
where the coefficient $C_{ij}^\alpha$ comprises the vertex coupling strengths
\begin{eqnarray}
C_{ij}^\alpha &=&
    \frac{7}{3} \kappa^2  \tan^2\beta \cos^2 2\beta \,
  y_{e_i}^{\rm SM} g_i^\alpha y_{e_j}^{\rm SM} g_j^\alpha, 
\end{eqnarray}
and the loop integral is represented by function $F$,
\begin{multline}
\!\!\!\!\! F(m_{H^\pm}^{},m_\Phi^{}, m_\chi, m_{\Sigma_{\alpha}}) =
 \left(\frac{1}{16\pi^2}\right)^3 \frac{(-m_{\Sigma_{\alpha}}^{})}{m_{\Sigma_{\alpha}}^2-m_\chi^2} \,
 \frac{v^2}{m_{H^\pm}^4}  \\
  \times \int_0^{\infty} \!\! x dx
  \left\{B_1(-x,m_{H^\pm}^{},m_\Phi^{})-B_1(-x,0,m_\Phi^{})\right\}^2
  \left(\frac{m_{\Sigma_{\alpha}}^2}{x+m_{\Sigma_{\alpha}}^2}-\frac{m_\chi^2}{x+m_\chi^2}\right)\nonumber \, ,
\end{multline}
where $B_1$ denotes the Passarino-Veltman function
for one-loop integrals~\cite{passarino-veltman}.
This produces small neutrino masses
with non-suppressed couplings: the values of $\mathcal{O}(1)$ for the quartic $\kappa$ and the appropriate Yukawa couplings easily reproduce neutrino masses $m_\nu \sim 0.1$ eV.

%%%%%%%%%%%%%%%%%%%%%%%%%%%%%%%%%%%%%%%%%%%%%%%%%%%%%%%%%%%%%%
\subsection{Vacuum stability and perturbativity} 
%%%%%%%%%%%%%%%%%%%%%%%%%%%%%%%%%%%%%%%%%%%%%%%%%%%%%%%%%%%%%%

It is in order to address the UV behavior of these two complementary loop scenarios. 
A summary of the detailed outcome of the minimal one-loop scenario with a Higgs partner from extra adjoint representation is presented in the first row in Table~\ref{Final}. It is contrasted to a more baroque three-loop model based on a Higgs partner from 2HDM sector in the second row in Table~\ref{Final}.
\begin{table}[h!]
\footnotesize
$$\begin{array}{lc|c|c|c||c|c|c|c|c|}
\hbox{Model}&\hbox{$J^{CP}_{H}$} & \hbox{$\Gamma_{H}$}&\hbox{Production}  & \hbox{Landau Pole}  &\hbox{Br$_{WW}$}& \hbox{Br$_{\gamma\gamma}$}&\hbox{Br$_{Z\gamma}$}&\hbox{Br$_{ZZ}$}&\hbox{Br$_{t\bar{t}}$}\\
\hline
\rowcolor[cmyk]{0.1,0,0.1,0}
\multicolumn{1}{c}{\cellcolor[cmyk]{0,0,0.2,0} \hbox{1-loop}}  &0^{++} &\hbox{3 GeV} & \hbox{$\gamma\gamma$-fusion} &\hbox{Absent} &64\%\ & 7\%\ &6\%\ &23\%\ &- \\
\rowcolor[cmyk]{0.2,0.0,0,0.0}
\multicolumn{1}{c}{\cellcolor[cmyk]{0,0.2,0,0} \hbox{3-loop}} &0^{++}&\hbox{44 GeV}&\hbox{$gg$-fusion}& 10^6 \ \hbox{GeV} &26\%\ & 1\%\ &4\%\ &7\%\ &61\%\ \\
\end{array}
$$
\caption{ Comparison between the neutrino mass models for $m_H=700$ GeV presented in~\cite{Antipin:2017wiz}. 
}
\label{Final}
\end{table}
Notably, the three-loop model is under a well known threat that invoking large multiplets~\cite{Cirelli:2005uq} leads to Landau poles (LP) considerably below the Planck scale.
For the $SU(2)_L$ gauge coupling $g_2$, this threat has been addressed in~\cite{Sierra:2016qfa}. An exposure to additional scrutiny presented in  \cite{Antipin:2017wiz} shows that this LP appears around $10^6$ GeV.
As for the quartic couplings, the focus there has been on the ``mixed" scalar couplings which might
lead to observable diphoton signal, and the negative values for some of them might
endanger the stability of the scalar potential and perturbative control over the model. These quartic couplings can be read from two scalar potentials contained in~(\ref{scalarpot}):
\begin{eqnarray}\label{scalarpot-chi-Phi}
  V_m(H_{\bf{1}},H_{\bf{2}},\chi) &\supset& (\tau_1H_{\bf{1}}^\dagger H_{\bf{1}} 
  +  \tau_2 H_{\bf{2}}^\dagger H_{\bf{2}}) \chi^\dagger \chi \ ,  \\
 \nonumber 
V_m(H_{\bf{1}},H_{\bf{2}},\Phi) &\supset&  (\sigma_1H_{\bf{1}}^\dagger H_{\bf{1}}+\sigma_2H_{\bf{2}}^\dagger H_{\bf{2}}) \Phi^\dagger \Phi +  (\sigma'_1H_{\bf{1}}^* H_{\bf{1}}+ \sigma'_2H_{\bf{2}}^* H_{\bf{2}}) \Phi^* \Phi . 
\end{eqnarray}
From the the dominant contribution to the 1-loop beta functions of the mixed quartics $x=\tau_{1,2}$ and $y=\sigma_{1,2}$ given in \cite{Hamada:2015bra}:
\begin{equation}
\beta_{x} \sim 4 x^2 - \frac{153}{2} x g_2^2 + 36 g_2^4 \;, \qquad \beta_y \sim 4 y^2 - \frac{81}{2} y g_2^2 + 18 g_2^4 \;,
\end{equation}
the mixed quartics develop LP $\sim 10^6$ GeV together with the $g_2$ coupling.

The appearance of $\sim 10^6\GeV$ Landau pole (LP) for the $\SU{2}_L$ gauge coupling $g_2$ \cite{Sierra:2016qfa} eliminates the three-loop model from a unification framework. In contrast, as demonstrated in~\cite{Antipin:2017wiz}, the one-loop model which we will call the BPR model, exhibits, in addition to the absence of LP, the perturbativity and stability up to the Planck scale.
Moreover, the above requirements with respect to Yukawa and quartic couplings may remain valid also when including extra color octet or color sextet scalar fields~\cite{Heikinheimo:2017nth}.
As shown in~\cite{Kumericki:2017sfc}, adding these fields may be crucial to achieve the proper gauge unification.
\begin{table}[th!]
\caption{Contributions of BPR states to RGE running, with
threshold weights $r_{k}$ defined in Eq.~\eqref{eq:rk}. From Sec. 4on the weak hypercharge is normalized by $Q = T_3 + Y$.}
\label{tab:BPR-beta-coefficents}
\centering
\renewcommand{\arraystretch}{1.2}
\begin{tabular}{c c| c c}
 \multicolumn{2}{c|}{$k$} &  $\Delta B_{23}$  & $\Delta B_{12}$   \\ \hline
$h^+$ & $(1,1,1)$ & $0$& $\hphantom{+}\frac{1}{5} r_k$ \\
$\Delta$ & $(1,3,0)$ & $\hphantom{+}\frac{1}{3} r_k$& $- \frac{1}{3} r_k$ \\
$E_{L,R}$ & $(1,2,-1/2)$ & $\hphantom{+}\frac{1}{3} r_k$& $- \frac{2}{15} r_k$ \\
\end{tabular}
\end{table}

\vspace{0.5cm}

%%%%%%%%%%%%%%%%%%%%%%%%%%%%%%%%%%%%%%%%%%%%%%
\section{Renormalizable \SU{5}  Gauge Unification}%%%%%%%%%%%%%%%%%%%%%%%%%
%%%%%%%%%%%%%%%%%%%%%%%%%%%%%%%%%%%%%%%%%%%%%%

Let us recall how new fields from neutrino-mass models have been employed to the gauge coupling unification hinted first within the \SU{5} GUT. After realizing that there is no single gauge coupling crossing in the simplest \SU{5}  GUT, it was noticed that augmenting the SM  by a second Higgs
doublet and the corresponding supersymmetric (SUSY) partners enables a successful unification~\cite{Amaldi:1991cn} in minimal SUSY SM (MSSM). A decisive role~\cite{Li:2003zh} of the incomplete ({\em split}) irreducible representations ({\em irreps}) \irrepsub{5}{H} in the MSSM unification success, motivated numerous non-SUSY attempts to cure the crossing problem
with just six copies of the SM Higgs doublet field.

First studies of unification in the context of non-SUSY \SU{5} GUTs employed incomplete \SU{5} irreps corresponding to states belonging to tree-level seesaw models. The studies
in~\cite{Bajc:2006ia,Bajc:2007zf,Dorsner:2006fx} employed 
adjoint \SU{5} representation \irrepsub{24}{F} which contains both the fermion singlet and the $\TeV$-scale fermion triplet fields providing a low scale hybrid of type-I and type-III seesaw models. Similarly, Refs.~\cite{Dorsner:2005ii,Dorsner:2005fq,Dorsner:2007fy} employed \irrepsub{15}{S} \SU{5} representation with the $\TeV$-scale complex scalar triplet, employed in the type-II seesaw mechanism. The genuine one-loop model proposed by Zee~\cite{Zee:1980ai} has introduced only new \emph{scalar} fields, the charged singlet and the second complex doublet.
An embedding of original Zee model in renormalizable non-SUSY  \SU{5} setup has been studied in~\cite{Perez:2016qbo}.
Here we turn to BPR~\cite{Brdar:2013iea}  variant of Zee model
which offers additional vectorlike doublet and the scalar triplet as the largest weak representation. 
\begin{table}[th!]
\caption{BSM contributions to RGE running in the simplest \SU{5} embedding of the BPR mechanism. 
$H$ stands for SM Higgs doublet whose contribution has already been accounted
for by $b_i^{(SM)}$. The massless scalar leptoquarks $X$ and $Y$ get
absorbed into longitudinal components of massive gauge bosons. The $\beta$ coefficients of these scalars thus enter at the same scale as heavy
vectors (\emph{i. e.} $r_k \approx 0$).}
\label{tab:5_H-model-beta-coefficents}
\centering
\renewcommand{\arraystretch}{1.2}
\begin{tabular}{c c c| c c}
 \multicolumn{3}{c|}{$k$}  &  $\Delta B_{23}$  & $\Delta B_{12}$ \\
\hline
$H$ & $(1,2,1/2)$ & \multirow{ 2 }{*}{ \irrepsub{5}{H} }& $\hphantom{+}\frac{1}{6} r_k$& $- \frac{1}{15} r_k$ \\
$S_1$ & $(3,1,-1/3)$ & & $- \frac{1}{6} r_k$& $\hphantom{+}\frac{1}{15} r_k$ \\
\hline
$h^+$ & $(1,1,1)$ & \multirow{ 3 }{*}{ \irrepsub{10}{S} }& $0$& $\hphantom{+}\frac{1}{5} r_k$ \\
 & $(\bar{3},1,-2/3)$ & & $- \frac{1}{6} r_k$& $\hphantom{+}\frac{4}{15} r_k$ \\
 & $(3,2,1/6)$ & & $\hphantom{+}\frac{1}{6} r_k$& $- \frac{7}{15} r_k$ \\
\hline
 & $(1,1,0)$ & \multirow{ 5 }{*}{ \irrepsub{24}{S} }& $0$& $0$ \\
$\Delta$ & $(1,3,0)$ & & $\hphantom{+}\frac{1}{3} r_k$& $- \frac{1}{3} r_k$ \\
 & $(8,1,0)$ & & $- \frac{1}{2} r_k$& $0$ \\
$X, Y$ & $(3,2,-5/6)$ & & $\hphantom{+}\frac{1}{12} r_k$& $\hphantom{+}\frac{1}{6} r_k$ \\
$\bar{X}, \bar{Y}$ & $(3,2,5/6)$ & & $\hphantom{+}\frac{1}{12} r_k$& $\hphantom{+}\frac{1}{6} r_k$ \\
\hline
$E_{L,R} $ & $(1,2,-1/2)$ & \multirow{ 2 }{*}{ \irrepbarsub{5}{F} }& $\hphantom{+}\frac{1}{3} r_k$& $- \frac{2}{15} r_k$ \\
$$ & $(\bar{3},1,1/3)$ & & $- \frac{1}{3} r_k$& $\hphantom{+}\frac{2}{15} r_k$ \\
\end{tabular}
\end{table}
Notably, new scalars in BPR model, Table~\ref{tab:BPR-beta-coefficents}, when summed, mimick the two Higgs doublets which led to the MSSM unification success. In the following we review the embedding of these states in the \SU{5} setup, performed in~\cite{Kumericki:2017sfc}.
\begin{table}[th!]
   \caption{Contributions of \irrepsub{45}{H} to running. For a complete model
  the multiplets from Table~\ref{tab:5_H-model-beta-coefficents} are to be
   added. The states $S_{1}$, $S_3$ and $\tilde{S}_1$ are leptoquarks that,
   if light, would induce too fast proton decay.
   }
\label{tab:45_H-model-beta-coefficents}
\centering
\renewcommand{\arraystretch}{1.2}
\begin{tabular}{c c c| c c}
 \multicolumn{3}{c|}{$k$}  &  $\Delta B_{23}$  & $\Delta B_{12}$ \\
\hline
$\Sigma_a$ & $(1,2,1/2)$ & \multirow{ 7 }{*}{ \irrepsub{45}{H} }& $\hphantom{+}\frac{1}{6} r_k$& $- \frac{1}{15} r_k$ \\
$S_1\equiv\Sigma_b$ & $(3,1,-1/3)$ & & $- \frac{1}{6} r_k$& $\hphantom{+}\frac{1}{15} r_k$ \\
$S_3\equiv\Sigma_c$ & $(3,3,-1/3)$ & & $\hphantom{+}\frac{3}{2} r_k$& $- \frac{9}{5} r_k$ \\
$\tilde{S}_1\equiv\Sigma_d$ & $(\bar{3},1,4/3)$ & & $- \frac{1}{6} r_k$& $\hphantom{+}\frac{16}{15} r_k$ \\
$\Sigma_e$ & $(\bar{3},2,-7/6)$ & & $\hphantom{+}\frac{1}{6} r_k$& $\hphantom{+}\frac{17}{15} r_k$ \\
$\Sigma_f$ & $(\bar{6},1,-1/3)$ & & $- \frac{5}{6} r_k$& $\hphantom{+}\frac{2}{15} r_k$ \\
$\Sigma_g$ & $(8,2,1/2)$ & & $- \frac{2}{3} r_k$& $- \frac{8}{15} r_k$ \\
\end{tabular}
\end{table}

\subsection{BSM states restricted to BPR set}

The embedding of BPR states to the lowest respective \SU{5} multiplets has been a starting point in~\cite{Kumericki:2017sfc} to improve the gauge unification. The accomplishment of this attempt is controlled by the SM gauge coupling $\beta$ coefficients
\begin{equation}
    B_i = b_i^{\rm(SM)} + \sum_{m_k<\MGUT} \Delta b_i^{(k)} {r_k} \;,
    \label{eq:Bi}
\end{equation}
where the threshold weight factor of BSM state $k$ is defined as
\begin{equation}
    r_{k}  =  \frac{ \ln \MGUT/m_k }{\ln \MGUT /m_{Z}} \;.
    \label{eq:rk}
\end{equation}
Then, the unification condition $\alpha_1(M_{\rm GUT}) = \alpha_2(M_{\rm GUT}) = \alpha_3(M_{\rm GUT}) \equiv \alpha_{\rm GUT}$, expressed in the form of the B-test~\cite{Giveon:1991zm,Dorsner:2017ufx},
\begin{align}
\frac{B_{23}}{B_{12}} & \equiv \frac{B_2-B_3}{B_1-B_2}=\frac{\alpha_2^{-1}(m_Z)-\alpha_3^{-1}(m_Z)}{\alpha_1^{-1}(m_Z)-\alpha_2^{-1}(m_Z)}=\tfrac{5}{8}\,\frac{\sin^2\theta_w(m_Z)-\tfrac{\alpha_{EM}(m_Z)}{\alpha_3(m_Z)}}{\tfrac{3}{8}-\sin^2\theta_w(m_Z)}=0.718 \;, 
\label{eq:Btest}
\end{align}
yields the numerical value $0.718$ for the constants at $m_Z$ scale~\cite{Patrignani:2016xqp}, different from the SM value $0.528$. 
Also, since the expression for the GUT scale 
\begin{align}
\MGUT & = m_Z \, \exp\left (\frac{2\pi (\alpha_1^{-1}(m_Z)-\alpha_2^{-1}(m_Z))}{B_1-B_2}\right ) = m_Z  \, \exp\left (\frac{184.87}{B_{12}}\right ) 
\label{eq:defMGUT}
\end{align}
yields the value $\MGUT = 10^{13}\GeV$, additional BSM states are needed to
increase the unification scale up to at least $5\times 10^{15}\GeV$, consistent
with the proton lifetime bounds \cite{Miura:2016krn}.

Let us notice the BSM states from the one-loop neutrino mass model at hand  came as if on cue for an alternative non-super-symmetric GUT attempt. Namely, the additional scalars shown in Table~\ref{tab:BPR-beta-coefficents}, when summed, produce the effect of the vector-like fermion doublet from the last raw in this table.
Consequently, taking all BPR states close
to the electroweak scale (weight factors from Eq.~\eqref{eq:rk} being 
$r_k \sim 1$), the SM B-test value $0.528$ increases to $B_{23}/B_{12} = 0.974$, 
overshooting the required value $0.718$ from Eq.~\eqref{eq:Btest}.
This stems from three copies of BPR vector-like
lepton doublets, which actually double the RGE effect of previously mentioned six Higgs doublets. One might achieve the correct crossing of couplings if vector-like leptons in
Table~\ref{tab:5_H-model-beta-coefficents}
are set at the intermediate scale, with the factor $r_k \sim 0.5$.
Still, as shown in~\cite{Kumericki:2017sfc}, with
the simplest possible \SU{5} embedding of the SM Higgs doublet in the \irrepsub{5}{H}, the unification scale would be too low as long as we use only BPR states.
Therefore, Ref.~\cite{Kumericki:2017sfc} opened \irrepsub{45}{H} or 
\irrepsub{70}{H}  which contain both the SM Higgs doublet and the additional BSM states.

\subsection{Viable unification scenarios}

Since the unification with the Standard Model (SM) Higgs doublet restricted to  irrep \irrepsub{5}{H} failed, the next attempt in \cite{Kumericki:2017sfc} proceeded with the SM Higgs in a mixture of \irrepsub{5}{H} and \irrepsub{45}{H}. As a bonus,
the \irrepsub{45}{H} irrep enables to account for the GUT fermion mass relations by implementing the Georgi-Jarlskog mechanism~\cite{Georgi:1979df}. 
 The $\beta$ coefficients of the extra states from scalar \irrepsub{45}{H} can be found in Table~\ref{tab:45_H-model-beta-coefficents}, which should
be added to states in Table~\ref{tab:5_H-model-beta-coefficents} to
obtain a complete embedding of the SM Higgs and the BPR states into \SU{5} multiplets.

This opens immense unification possibilities which have been restricted by imposing a set of plausible criteria detailed in~\cite{Kumericki:2017sfc}. A search algorithm developed there
has selected seven successful scenarios listed in Table \ref{tab:H45}.
\begin{table}
\renewcommand{\arraystretch}{1.2}
\caption{\label{tab:H45} Seven unification scenarios with SM Higgs in
${\mathbf{5}}$ and/or ${\mathbf{45}}$ of SU(5) and BPR states fixed at $\sim 0.5\TeV$.
  }
\begin{center}
\begin{tabular}{cc|ccccccc}
    \multicolumn{2}{c|}{irreps} & \multicolumn{7}{c}{$m_{k}$ [TeV]} \\
    SM & SU(5) &  A1  &  A2  & A3  & A4  &  A5  & A6  & A7 \\ \hline
    $(3,1, 1/3)$ & \irrepbarsub{5}{F} &  5000  &   & $2.3\times 10^6$  &
    450  &    & $2\times 10^5$  &  \\ \hline
    $(\bar{3},1,-2/3)$ & \multirow{2}{*}{\irrepsub{10}{S}} &  &  &  &   &  
    &      & 2.4 \\
    $(3,2, 1/6)$ &  & 0.5 &  & 0.5 & 0.5  &  &  0.5 & 0.5  \\ \hline
    $(8,1,0)$ & \irrepsub{24}{S} &  & 0.5 & 0.5 & & 0.5 & 0.5 &   \\ \hline
    $(1,2, 1/2)$ & \multirow{3}{*}{\irrepsub{45}{H}} &  &  &  & 0.5 &
     260 & 0.5  &          \\
    $(\bar{6},1, -1/3)$ & &  & 90  &  &  & 0.5 &  & 0.5\\ 
    $(8,2, 1/2)$  & & 0.5 & 0.5 & 0.5 & 0.5 & 0.5 & 0.5 & 0.5   \\ \hline
    \multicolumn{2}{c|}{$\MGUT^{\rm max}/(10^{15}\,\GeV)$} & 2.8 & 2.5 & 6.2 &
     2.8 & 2.8 & 6.2 & 6.5 
\end{tabular}
\end{center}
\end{table}
In all of them, a light colored scalar $(8,2,1/2)$  provided by \irrepsub{45}{H} plays a decisive role. 

For the alternative choice of the SM Higgs belonging to \irrepsub{70}{H} instead
to \irrepsub{45}{H}, our search algorithm selects four scenarios (A1, A3, A4 and A6) from  Table \protect\ref{tab:H45}, and allows for three additional scenarios presented in~\cite{Kumericki:2017sfc}.
Thereby, we have explicitly excluded representation $(15, 1, -1/3)$
which leads to Landau poles below $\MGUT$, unless it is
heavier than $10^7\,\GeV$
\cite{Heikinheimo:2017nth} so that its effect on the RG running would be diminished.

Notably, in these additional scenarios the
BPR vector-like leptons are assigned to complete irrep \irrepsub{5}{F}, that
do not affect the RGE running. 
Since in these latter scenarios only the scalar \SU{5} irreps are incomplete, 
an eventual verification of them would be in support of a conjecture in~\cite{Cox:2016epl} which is  based on embedding SM Higgs into \irrepsub{5}{H}, that such splitting could be an inherent feature of the scalar multiplets.

To conclude, in our procedure of renormalizable \SU{5} embedding, the colorless BPR particles employed in the neutrino mass model get accompanied by the colored partners to enable a successful unification. 
We decide to keep sufficiently heavy those among the colored leptoquark scalars which present a threat to proton stability, and the other colored states may  
play a model-monitoring role both through the LHC
phenomenology~\cite{Giveon:1991zm,Dorsner:2017ufx} and through tests at
Super(and future Hyper)-Kamiokande \cite{Miura:2016krn} experiments.

We also point out that in most of the allowed parameter space the color octet scalar $(8,2,1/2)$ is the most promising BSM state for the LHC searches, and
as such is studied already in \cite{Perez:2016qbo}. 
Additional colored states in the specific gauge unification scenarios in Table~\ref{tab:H45} call for a study of characteristic exotic signals at the LHC, in order to make these specific models falsifiable.

%%%%%%%%%%%%%%%%%%%%%%%%%%%%%%%%%%%%%%%%%%%%%%
\section{Further Directions}%%%%%%%%%%%%%%%%%%%%%%%%%
%%%%%%%%%%%%%%%%%%%%%%%%%%%%%%%%%%%%%%%%%%%%%%

The need for neutrino masses modifies minimal ``old'' SM to some new ``$\nu$SM'', where, in order to understand the lightness of active neutrinos, one commonly introduces
new heavy degrees of freedom. While no evidence for such new particles has been found so far at the LHC, evidences from astrophysics and cosmology point at 23$\%$ of the energy density of the Universe provided by DM candidates such as heavy neutrino-like weakly interacting massive particles (WIMPs).

Bearing in mind an accidental stability of our ordinary matter, it is possible that a DM setup is also accidentally stable.
The so-called minimal dark matter model (MDM)~\cite{Cirelli:2005uq} with single aDM particle is probably the simplest such scenario. In this setup the stability can be guaranteed for a neutral component of a large enough fermion or scalar $SU(2)_L$ multiplet, which can not form $SU(2)_L$ invariant renormalizable (or dimension-five) interaction-terms with the SM multiplets. Since the neutral (DM) component cannot have the tree-level interactions to the Z boson, both the hypercharge $Y$ and the isospin-component $T_3$ must vanish. This is possible only for odd multiplets: fermion quintuplet or higher, and scalar septuplet or higher. When employing the fermion quintuplets as seesaw mediators, like in two models presented at the beginning, they are in conjunction with appropriate scalar quadruplets. 
This destroys the stability of MDM Majorana quintuplet from Sec. 2.2, unless additional protecting symmetry is assumed. By imposing the discrete  $Z_2$ symmetry, the mass of $\Sigma^0$, as the DM particle, is fixed by the relic abundance to the value~\cite{Cirelli:2005uq} $M_\Sigma \approx 10\ \rm{TeV}$. In Sec. 2.2  the loop contribution predominates, and the choice $\lambda_5=10^{-7}$ gives enough suppression to lead to small neutrino masses with large Yukawas, $Y \sim 0.1$. In this part of the parameter space the model could have interesting LFV effects studied in~\cite{Cai:2011qr}. 

However, high-order SM multiplets are not compatible with the gauge unification. Lower multiplets belonging to a studied  non-scotogenic variant of the BPR model have resulted in a list of viable unification scenarios. Thereby, the split   \SU{5} multiplets are inevitable. The deduced particle sets which may provide gauge unification generalize the fine-tuned anomaly-free set of ordinary SM particles. It is tempting to impose a DM mission on the one-loop model at hand. 
In fact, its scotogenic variant has been in focus of the BPR paper~\cite{Brdar:2013iea} on the trace of the original scotogenic model~\cite{Ma:2006km} followed by the new scotogenic model~\cite{Ma:2013yga} where, an othervise imposed DM protecting symmetry, would be a remnant of an underlying gauge symmetry.
The program of gauge unification would acquire entirely new form in the context of DM species providing new force carriers.

To conclude, we adopted a non-supersymmetric GUT context devoid of naturalness as a guiding principle for an UV completion. Instead, we rely on incomplete multiplets as a generic feature of the scalar sector, related to gauge unification.
Regarding the uniqueness of the unification solution, it would be welcome to possess a monitoring
principle controlling possible UV completions. In fact, a candidate for such principle has been in focus of Hvar 2018 workshop, known under the name of "asymptotic safety"~\cite{Weinberg:1980}. It spells yet unconfirmed hope that quantum-gravity fluctuations may induce the asymptotically safe fixed point, controlling UV completions of the SM~\cite{Pelaggi:2017abg} with or without DM~\cite{Eichhorn:2018yfc} particle species.

\section*{Acknowledgments}
This work is supported by the Croatian Science Foundation project number 4418.

\end{document}